\documentclass[12pt]{article}
\usepackage{amsfonts,amsmath,epsf}

\def\bH {\mathbb{H}}
\def\bR {\mathbb{R}}

\newcommand{\vev}[1]{{\left< {#1} \right>}}

\newcommand{\Tr}{{\rm Tr\,}}

\newcommand{\cC}{{\cal C}}
\newcommand{\cD}{{\cal D}}

\newcommand{\cN}{{\cal N}}
\newcommand{\cO}{{\cal O}}
\newcommand{\cP}{{\cal P}}

\newcommand{\preprint}[1]{\begin{table}[t]  
             \begin{flushright}               
             {#1}                             
             \end{flushright}                 
             \end{table}}                     
\renewcommand{\title}[1]{\vbox{\center\LARGE{#1}}\vspace{5mm}}
\renewcommand{\author}[1]{\vbox{\center#1}\vspace{5mm}}
\newcommand{\address}[1]{\vbox{\center\em#1}}
\newcommand{\email}[1]{\vbox{\center\tt#1}\vspace{5mm}}

\begin{document}

\begin{titlepage}
\preprint{NSF-KITP-05-115\\
hep-th/0512150}

\title{Circular loop operators in conformal field theories}

\author{Nadav Drukker$^1$ and Shoichi Kawamoto$^2$}

\address{$^1$The Niels Bohr Institute, Copenhagen University\\
Blegdamsvej 17, DK-2100 Copenhagen, Denmark\\
\medskip
$^2$Rudolf Peierls Centre for Theoretical Physics,
University of Oxford,\\
1 Keble Road, Oxford, OX1 3NP, United Kingdom
}

\email{drukker@nbi.dk, kawamoto@thphys.ox.ac.uk}

\abstract{
We use the conformal group to study non-local operators in 
conformal field theories. A plane or a sphere (of any dimension) 
is mapped to itself by some subgroup of the conformal group, 
hence operators confined to that submanifold may be classified 
in representations of this subgroup. For local operators this 
gives the usual definition of conformal dimension and spin, but 
some conformal field theories contain interesting nonlocal operators, 
like Wilson or 't Hooft loops. We apply those ideas to Wilson 
loops in four-dimensional CFTs and show how they can be 
chosen to be in fixed representations of $SL(2,\bR)\times SO(3)$.
}

\end{titlepage}

Conformal field theories (CFTs) play an important role in 
physics. They arise naturally at the fixed points of the renormalization 
group flow and describe critical phenomena in a wide class of 
systems. From the theorists perspective CFTs offer an enlargement of the 
space-time symmetry group that puts constraints on the theory and 
makes it easier to study than a general quantum field theory.

Many field theories contain non-local observables, like Wilson loop 
operators, or topological defects, like Nielsen-Olesen vortices, or 
't Hooft loops. Such objects may appear also in theories that 
have a conformal symmetry, for example Wilson loops in $\cN=4$ 
supersymmetric Yang-Mills theory. 
In this note we propose some tools to study non-local operators 
in conformal field theories.

Let us recall the construction of local conformal operators. The 
conformal group in $d$-dimensional Euclidean space is $SO(d+1,1)$. 
The subgroup that will keep a fixed point (the origin) invariant is 
$SO(d)\times\bR$, comprising of rotations and the dilatation. Hence 
local operators may be classified by representations 
of this subgroup, the spin and conformal dimension.

To generalize this construction for non-local operators consider 
an $n$-dimensional sphere 
in $\bR^d$. The subgroup of the conformal group 
that maps the sphere to itself is $SO(d-n)\times SO(n+1,1)$. 
A simple way to see this symmetry is to map the sphere to a plane 
by a stereographic projection. $SO(n+1,1)$ is the conformal symmetry 
in this plane and $SO(d-n)$ are the rotations around the plane.
In the specific case of $n=1$ and $d=4$ on which we concentrate later 
we call these operators ``circular loop operators'' and the symmetry 
group is $SL(2,\bR)\times SO(3)$~\cite{kapustin}.

Our main proposition concerns any non-local operator localized on 
a sphere. The claim is
\begin{quote}
Operators localized on $S^n$ in a CFT can be classified by representations  
of $SO(n+1,1)\times SO(d-n)$ in much the same way that local 
operators are classified by spin and conformal dimension.
\end{quote}
This statement follows immediately from the preceding discussion. 
In the remainder of the note we will develop some tools for analyzing 
loop operators in this setting and demonstrate it in a few examples.

\subsubsection*{Symmetry}

An arbitrary CFT posses an $SO(d+1,1)$ symmetry generated by 
translations $P_\mu$, Lorentz transformations $M_{\mu\nu}$, 
the dilatation $D$ and special conformal transformations $K_\mu$. 
Those may be realized on the fields by \cite{Mack:1969rr}
\begin{equation}
\begin{aligned}[]
[P_\mu,\,\Phi(x)]=&-i\partial_\mu\Phi(x)\,,\\
[M_{\mu\nu},\,\Phi(x)]
=&-i(x_\mu\partial_\nu - x_\nu \partial_\mu
+\Sigma_{\mu\nu})\Phi(x)\,,\\
[D,\,\Phi(x)] =& -i(\Delta + x_\mu\partial_\mu)\Phi(x)\,,\\
[K_\mu,\,\Phi(x)]
=&-i(x^2\partial_\mu - 2x_\mu x^\nu \partial_\nu-2x_\mu\Delta
+2x^\nu \Sigma_{\nu\mu})\Phi(x)\,.
\label{diff-operators}
\end{aligned}
\end{equation}
$\Sigma_{\mu\nu}$ are Lorentz matrices acting on non-scalar fields.

Of those generators, the ones that leave the origin invariant are 
the Lorentz transformations and the dilatation which generate the 
subgroup $SO(d)\times\bR$. Once we classify local operators by 
representations of this subgroup we can construct the full 
multiplet by considering the action of the other generators, 
$P_\mu$ and $K_\mu$ on the local operators. Since those carry 
mass dimension $+1$ and $-1$ they can be regarded as raising 
and lowering operators, and if a state is annihilate by all the $K_\mu$ 
it's called a highest-weight state, or a primary operator.

Instead we are interested in the subgroup that maps a sphere $S^n$ of
radius~$R$ given by $\sum_{i=1}^{n+1} (x^i)^2=R^2$ 
to itself. This is generated by
\begin{align}
&  J_i = \dfrac{R}{2}P_i + \dfrac{1}{2R}K_i 
\qquad\hbox{and}\qquad
M_{ij}
& \hbox{for}\quad
i,j=1, \cdots , n+1\,,
\\
&  L_{i'}=\dfrac{R}{2}P_{i'}-\dfrac{1}{2R}K_{i'}
\qquad\hbox{and}\qquad
M_{i'j'}
& \hbox{for}\quad
i',j'=n+2, \cdots , d\,.
\label{subgroup}
\end{align}
The first operators $\{ J_i, M_{ij} \}$ generate 
$SO(n+1,1)$ while the second set of operators 
are the generators of $SO(d-n)$. 
One can easily check using the 
representation (\ref{diff-operators}) that the $SO(d-n)$ operators 
map every point on the sphere
to itself, while 
when restricted to the sphere the $M_{ij}$ and $J_i$ only include
the derivatives tangent to the surface of the sphere
and hence map the sphere to itself.

A simple way to realize this symmetry is by writing $\bR^d$ in a 
special coordinate system. Staring with $(\eta,\Omega_{n})$ as polar 
coordinate in $\bR^{n+1}$  and $(\zeta,\Omega_{d-n-2})$ in the 
remaining space, we define $\rho$ and $\theta$ by
\begin{equation}
\sin\theta=\frac{\zeta}{\tilde r}\,,
\qquad
\sinh\rho=\frac{\eta}{\tilde r}\,,
\qquad
\tilde r=\frac{\sqrt{(\zeta^2+\eta^2-R^2)^2+4R^2\zeta^2}}{2R}
=\frac{R}{\cosh\rho-\cos\theta}\,.
\end{equation}
The flat space metric is then written as
\begin{equation}
\begin{aligned}
ds^2=&\,d\eta^2+\eta^2 d\Omega_n^2+d\zeta^2+\zeta^2d\Omega_{d-n-2}^2\\
=&\,
\tilde r^2
\left(d\rho^2+\sinh^2\rho\, d\Omega_n^2
+d\theta^2+\sin^2\theta\,d\Omega_{d-n-2}^2\right)\,.
\label{metric:AdSxS_generic}
\end{aligned}
\end{equation}
One can immediately see that after dividing by the conformal factor 
$\tilde r^2$ one gets the metric on $\bH_{n+1}\times S^{d-n-1}$, 
where $\bH_{n+1}$  is the $n+1$-dimensional hyperbolic plane, also 
known as Euclidean $AdS_{n+1}$. 
The sphere at $\eta=R$ and $\zeta=0$ is mapped to the boundary 
of $\bH_{n+1}$ at $\rho\to\infty$, and is therefore invariant under the 
isometries of this space $SO(n+1,1)\times SO(d-n)$.

Since the sphere is an orbit of the subgroup we may classify
operators constrained on the sphere by representations of this
subgroup.
There are $(n+2)(d-n)$ generators of the full group that are not in this
subgroup.
 Those transform in the $({\bf n+2},{\bf d-n})$ representation of 
$SO(n+1,1)\times SO(d-n)$.
To construct the operators explicitly 
we may start with the dilatation,
which commute with all the $M_{ij}$.
Acting on it with $J_i$ and $L_{i'}$ will give the 
remaining operators in the coset.

The coset generators map operators on the sphere in different
representations
to each other in much the same way that $P_\mu$ and $K_\mu$ 
related local conformal operators of different dimensions.

\subsubsection*{Example: Wilson loops in four dimensions}

Thus far we have considered operators on a sphere of arbitrary dimensions, 
In the rest of the paper we will concentrate on the
case of $d=4$ and $n=1$, so the symmetry is $SL(2,\bR)\times SO(3)$. 
For the $SL(2,R)$
generators we take $J_0=-M_{12}$ and $J_\pm = J_1 \pm i J_2$.
For concreteness we label the 
angular coordinate in the plane of the circle by $\psi$ and in the 
other two directions by $\phi$.

The specific example we focus on is of Wilson loop operators in 
four-dimensional $\cN=4$ supersymmetric Yang-Mills. Another example 
one may consider is classical electro-magnetism. From the group theory analysis 
we know that they can be classified by representations of 
$SL(2,\bR)\times SO(3)$.

Consider a Wilson loop along a circle of radius $R$
\begin{equation}
W=\Tr\cP\, e^{i\oint (A_\psi(R,\psi)+iR\Phi_6(R,\psi)) d\psi}\,.
\label{loop}
\end{equation}
This operator, with the inclusion of an extra adjoint scalar ($\Phi_6$) in the 
exponent, is a very natural observable in the 
supersymmetric theory \cite{ads-loops}.

At the classical level both this operator and the one without the scalar 
term will be in the trivial representation of 
$SL(2,\bR)$, but this may be modified by quantum corrections.
We expect the one with the scalar, which is supersymmetric, 
to remain in the 
trivial representation even after including quantum corrections.

If those were the only objects that can be studied by our classification it 
would be hardly worth the effort, but this is not the case. A general 
Wilson loop will not be circular, so it will not preserve the symmetry, but 
if the geometry is close enough to the circle, we may expand it about 
the circular one. Small variations in the shape of the loop may be 
replaced by local insertions into the loop (see also \cite{expand})
\begin{equation}
\frac{\delta}{\delta x^\mu(s)}W
\sim \Tr\cP\, iF_{\mu\nu}\dot x^\nu(s)e^{i\oint\cdots}\,.
\end{equation}
This operator is gauge invariant and circular, so may be classified in 
representations of $SL(2,\bR)$. 
In general, for any set of operators $\cO_i$ transforming in the adjoint 
representation of the gauge group we may consider the operator
\begin{equation}
W[\cO_1(\psi_1)\cdots \cO_k(\psi_k)]
=\Tr\cP\left[ \cO_1(\psi_1)\cdots\cO_k(\psi_k)
e^{i\oint (A_\psi(R,\psi)+iR\Phi_6(R,\psi)) d\psi}\right]\,.
\end{equation}
Those are the types of operators we suggest should be studied in this fashion.

\subsubsection*{Example: Smeared scalar operators}

Let us now turn to calculating the dimension, and focus on the simple 
case of a single scalar insertion. Define the Fourier 
components
\begin{equation}
W[\cO]^{(m)}
=\frac{1}{2\pi}\Tr\cP\,\int d\psi'\,\cO(\psi')\,e^{im\psi'}
e^{i\oint (A_\psi(R,\psi)+iR\Phi_6(R,\psi)) d\psi}\,.
\end{equation}
In perturbation theory the holonomy does not contribute at tree level, so 
in this case
$W[\cO]^{(m)}$ reduces to the smeared local operator
\begin{equation}
\cO^{(m)}
=\frac{1}{2\pi}\int d\psi\,\Tr\cO(\psi)\,e^{im\psi}\,.
\end{equation}

As stated, in the $\rho,\psi$ coordinate system, the $SL(2,\bR)$ symmetry 
is just the isometry of $\bH_2$. One can show that the action of the 
operators in (\ref{subgroup}) on scalar fields of dimension $\Delta$ is 
given by
\begin{equation}
\begin{aligned}[]
[J_0,\,\Tr\cO]=&\,i\partial_\psi\Tr\cO\,,\\
[J_\pm,\,\Tr\cO]=&\,e^{\pm i\psi}\tilde r^{-\Delta}
\left(-i\partial_\rho \pm \coth\rho\,\partial_\psi\right)
\tilde r^\Delta\Tr\cO\,.
\end{aligned}
\end{equation}
When the operator is along the circle we may take $\rho\to\infty$ 
hence the action of $J_\pm$ simplifies to
\begin{equation}
[J_\pm,\,\Tr\cO]=\,e^{\pm i\psi}
\left(i\Delta \pm \partial_\psi\right)\Tr\cO\,.
\end{equation}

Acting directly on the operators one finds after integration by parts
\begin{equation}
[J_0,\,\cO^{(m)}]=m\cO^{(m)}\,,\qquad
[J_\pm,\,\cO^{(m)}]=
-i (1-\Delta\pm m)\cO^{(m\pm1)}\,.
\label{direct}
\end{equation}
The Casimir is $J^2=(J_+J_-+J_-J_+)/2-J_0^2=-\Delta(\Delta-1)$, 
which is consistent with representations with principle quantum
number\footnote{%
On the representation of $SL(2,\bR)$ and notations, see \cite{SL2-repr}.} 
$j=\Delta$ or $j=1-\Delta$. Since the representation includes all integer 
values of $m$ it's in the continuous series and since for integer 
$\Delta$ the lowering operator $J_-$ annihilates the operator with 
$m=1-\Delta$, it has $j=1-\Delta$. This representation is non-unitary 
and for integer $\Delta$ includes as sub-representations the states 
with $m>1-\Delta$, those with $m<\Delta-1$ or their intersection, 
which is a finite-dimensional representation (the same as the unitary 
representations of $SU(2)$).

Instead of considering the Fourier modes of fields we can consider 
local insertions into the Wilson loop, say at $\psi=0$ and 
$\psi=\pi$. Now we can look at the generators
\begin{equation}
\tilde J_0=-i J_1\,,\qquad
\tilde J_\pm=iJ_0 \mp i J_2\,.
\end{equation}
A local primary operator of dimension $\Delta$ at $\psi=0$ will be 
annihilated by $\tilde J_-$ and have an eigenvalue $\Delta$ for 
$\tilde J_0$, acting with $\tilde J_+$ will give a tower
\begin{equation}
\Tr\cO(0)\,,\qquad
-2\,\Tr\partial_\psi\cO(0)\,,\qquad
4\,\Tr\partial_\psi^2\cO(0)-2\Delta\Tr\cO(0)\,,\qquad
\cdots
\end{equation}
with increasing values of $\tilde J_0$ 
that are in the discrete unitary representation $\cD_\Delta^+$. 
In a similar way a primary local operator at $\psi=\pi$ will have 
$\tilde J_0=-\Delta$ and will be the highest weight state in the 
conjugate representation $\cD_\Delta^-$. There are other operators 
that are in the continuous representations 
$\cC_{1-\Delta}^\alpha$ in this basis with arbitrary $\alpha$.

Note that it is possible to map flat $\bR^4$ to $S^3\times\bR$ by 
a conformal transformation that will map $\psi=0$ to past infinity 
and $\psi=\pi$ to future infinity. Then $\tilde J_0$ will 
correspond to time translation and the unitary representations 
will correspond to physical states in the Wick rotation of that space.

For local operators a simple way of calculating the dimension is 
through the two point function
\begin{equation}
\vev{\Tr\bar\cO(x)\Tr\cO(0)}\sim\frac{1}{x^{2\Delta}}\,.
\end{equation}
One can study the representations of circular loop operators in a 
similar fashion. 
Consider the two point function of a loop operator of radius $R$ and mode 
number $m$ and another or radius $\eta$ at $\zeta$ with an insertion at 
$\psi$
\begin{equation}
\vev{W[\bar\cO(\eta,\psi,\zeta)]\ W[\cO]^{(m)}}
\end{equation}
As stated, at tree level we may replace the Wilson loop with the smeared 
local operator $\cO^{(m)}$. If $\Tr\cO$ has dimension $\Delta$, 
the two point function is given by
\begin{equation}
\vev{W[\bar\cO(\eta,\psi,\zeta)]\ W[\cO]^{(m)}}
=\frac{1}{(2\pi)^{2\Delta+1}}\int d\psi'\,
\frac{e^{im\psi'}}{(\zeta^2+\eta^2+R^2-2R\eta
  \cos(\psi-\psi'))^{2\Delta}}\,.
\end{equation}
After writing this in terms of $\rho$ and $\theta$, this integral is
\begin{equation}
\frac{e^{im\psi}}{2\pi(8\pi^2R\tilde r)^\Delta}\int d\psi'\,
\frac{e^{im\psi'}}{(\cosh\rho-\sinh\rho\cos\psi')^{2\Delta}}
=\frac{e^{im\psi}}{(8\pi^2R\tilde r)^\Delta}
\frac{\Gamma(\Delta+|m|)}{\Gamma(\Delta)}
P_{-\Delta}^{-|m|}(\cosh\rho)\,,
\end{equation}
where $P_{-\Delta}^k$ is a version of the associated Legendre 
function which is defined on the positive real line but has a branch
cut along $(-\infty,1]$. It is equal to
\begin{equation}
P_{-\Delta}^k(\cosh\rho)=\left(\coth\frac{\rho}{2}\right)^k
{}_2F_1\left(\Delta,1-\Delta;1-k\,
\big| - \sinh^2\frac{\rho}{2}\right)\,.
\end{equation}
One can see that this function accompanied by the phase factor is an
eigenfunction of the Laplacean of $\bH_2$ with eigenvalue
$-\Delta(\Delta-1)$.

Acting with the generators $J_i$ on the coordinates $\rho$ and $\psi$ 
one finds the relations
\begin{equation}
\begin{aligned}
\vev{(J_0 W[\bar\cO(\eta,\psi,\zeta)])\ W[\cO]^{(-m)}}
=&\,m\vev{W[ \Bar{\cO}(\eta,\psi,\zeta)]\ W[\cO]^{(-m)}}
\\
\vev{(J_\pm W[\bar\cO(\eta,\psi,\zeta)])\ W[\cO]^{-(m\pm1)}}
=&\,-i(\Delta\pm m)
\vev{W[\Bar\cO (\eta,\psi,\zeta)]\ W[\cO]^{(-m)}}\,.
\end{aligned}
\end{equation}
Combining this result with the direct application of the symmetry 
generators on $W[\cO]^{(m)}$ in (\ref{direct}) we can verify the 
Ward identities
\begin{equation}
  J_i \vev{W[\Bar\cO(\eta,\psi,\zeta)]\, W[\cO]^{(m)}} =0\,,
\end{equation}
up to contact terms.

\subsubsection*{The OPE}

An important property of local operators in a CFT is the existence of 
the operator product expansion (OPE). It is possible to replace two nearby 
operators of dimensions $\Delta_1$, $\Delta_2$ with a series of 
operators of dimensions $\Delta_k$
\begin{equation}
\cO_2(x)\cO_1(0)
=\sum C_{21}^k x^{\Delta_k-\Delta_1-\Delta_2}\cO_k\,.
\label{OPE}
\end{equation}
One may hope that a similar property applies to loop operators. To 
justify that, note that loop operators specify boundary 
conditions on $\bH_2$ and if one considers two nearly coincident loops, 
they are both near the boundary. We may then look at the result in the 
bulk and write it in terms of one set of boundary conditions.

If our loop operators are made of smeared gauge invariant operators
on the circles at $\rho=\infty$ and $(\rho,\theta)$,
the OPE is inherited from that of the local operators
as
\begin{equation}
\begin{aligned}
\cO_2^{m_2}(\rho,\theta) \cO_1^{m_1}
=&
\sum_k C_{21}^k \cO_k^{m_1+m_2} (2R\tilde r)^{\delta/2}
\frac{\Gamma(-\delta/2 -m_2)}{\Gamma(-\delta/2)}
P_{+\delta/2}^{-|m_2|}(\cosh\rho)\,,
\label{composite}
\end{aligned}
\end{equation}
where $\delta=\Delta_k-\Delta_1-\Delta_2$.
In comparing to the OPE of local operators, the spatial dependence 
is more complicated and includes both the $\tilde r$ dependence as 
well as the associated Legendre function.

\subsubsection*{Descendants}

Similarly to local operators, also in the case of circular loop operators
different $SL(2,\bR)$ representations may be related to each other by
the action of generators in the coset $SO(5,1)/SL(2,\bR)\times SO(3)$. 
This coset is nine-dimensional with the operators transforming in 
the $({\bf 3},{\bf 3})$ of $SL(2,\bR)\times SO(3)$. One 
element in the coset is the dilatation operator $D$.
This operator commutes with both $J_0$ and $L_0=-M_{34}$ and it's easy to 
write down the other 8 generators by commuting it with the raising 
and lowering operators.

Acting with the dilatation operator on the Wilson loop (\ref{loop}) gives
\begin{equation}
W=R\,\Tr\cP\int d\psi(F_{\eta\psi}(\psi)+iRD_\eta\Phi(\Psi))
e^{i\oint (A_\psi+iR\Phi) d\psi'}\,.
\end{equation}
This operator is in the $j=-1$ representation of
$SL(2,\bR)$ since the Wilson loop itself is in the trivial representation.
Acting with $J_\pm$ on this operator will insert phase factors 
$e^{\pm i\psi}$ into the integral. The action of the $SO(3)$ generators 
will replace $\eta$ in the field-strength and derivative with the $3$
and $4$ directions. 

In general the action of $D$ on a circular operator in the representation 
with $j=1-\Delta$ will yield a reducible representation---the
product of that representations with $j=-1$, 
including states with $j=-\Delta$, $1-\Delta$ and $2-\Delta$. 
One can check this explicitly for our example by repeating the calculation in 
(\ref{direct}) on the Fourier modes of $[D,\,\Tr\cO]$. 
We will call a circular loop operator a primary if 
the result will include only the 
representation with $j=-\Delta$. Indeed if we consider an insertion 
$\cO$ into the loop such that $\Tr\cO$ is a primary local operator, 
the resulting loop operator will be a primary (at tree level) by this 
definition.

\subsubsection*{Outlook}

We have presented a method of classifying non-local operators in 
conformal field theories that are constrained to spherical or planar 
subspaces. This classification is based on the subgroup of the 
conformal group that preserves this subspace. We demonstrated 
this on the example of Wilson loops in four dimensions, which 
may be organized into representations of $SL(2,\bR)\times SO(3)$.

The simplest circular Wilson loop in $\cN=4$ supersymmetric 
Yang-Mills theory seems to be in the trivial representation. For 
other Wilson loops we suggested expanding nearly-circular operators 
in terms of circular loops with insertions into them. Those will 
fit into other representations of the symmetry group.

In the examples we studied we calculated the representations only 
at tree level and have so far not considered quantum corrections. 
The dimensions of local operators get corrected 
in perturbation theory, and so should the $SL(2,\bR)$ representations. 
Those may be calculated by looking at the two-point function of 
Wilson loops, as was done for the simplest two circles in 
\cite{Plefka:2001bu}. The result is not expected to 
remain Legendre functions with modified index, rather the two-point 
function would be a representation of the symmetry whose generators 
are modified by quantum corrections.

One may wish to study other objects, for example 't Hooft loops. 
Those may be described semiclassically by a magnetic flux in some 
$U(1)$ subgroup of the gauge group sourced along the circle. After 
the map to $\bH_2\times S^2$ those would correspond to a constant 
magnetic flux on $S^2$ \cite{kapustin}. Since this has no structure on 
$\bH_2$ it would seem natural to conjecture that this too is in the 
trivial representation of $SL(2,\bR)$.

Again, there should be generalizations corresponding to small 
deformations of the circular 't Hooft loop. It may be possible to 
study those by a similar semiclassical description, where now other 
components of the electromagnetic field would be excited. We 
leave the study of these objects to the future.

In the case of $\cN=4$ supersymmetric Yang-Mills theory the basic 
circular loop preserves half the supersymmetries and 
has some remarkable properties when calculated both in perturbation 
theory and in the dual string theory on $AdS_5\times S^5$
\cite{Erickson:2000af}. 
One consequence of the supersymmetry is that the 
group presented above is enlarged to a supergroup with 16 fermionic 
generators. This leads to a much richer structure that will be 
studied elsewhere. Furthermore, in that case one can look at the string 
theory duals of those operators described by 
classical string solutions in $AdS_5\times S^5$ and their $SL(2,\bR)$ 
representations should be calculable there too.

We have applied those ideas only to four-dimensional theories but 
the same could be done in arbitrary dimensions. In particular it 
would be interesting to study the spherical surface observable in the 
six-dimensional theory dual to $AdS_7\times S^4$.

\subsubsection*{Acknowledgments}

We would like to thank Niklas Beisert, Poul Henrik Damgaard, Shinji Hirano, 
Yaron Oz and Matthias Staudacher for helpful discussions. N.D. is 
grateful to Tel Aviv University, the Albert Einstein Institute in 
Golm and the KITP in Santa Barbara for their hospitality during the 
course of this work. 
S.K. thanks the Niels Bohr Institute where he carried out most of the
work as a post-doc. The work of N.D was supported in part by the National 
Science Foundation under grant No. PHY99-07949. The work of 
S.K. was supported by 
ENRAGE (European Network on Random Geometry), a Marie Curie
Research Training Network supported by the European Community's
Sixth Framework Programme, network contract MRTN-CT-2004-005616.


\end{document}